\documentclass[aps,final]{revtex4}%
\usepackage[english]{babel}
\usepackage{graphics,graphicx}
\usepackage{amssymb,amsfonts}
\usepackage{amsmath}
\usepackage{amsfonts}
\usepackage{amssymb}
\usepackage{graphicx}%

\begin{document}
\title{Cascade Birth of Universes in Multidimensional Spaces}
\author{S. G. Rubin}
\affiliation{Moscow Engineering Physics Institute, Kashirskoe sh. 31, Moscow, 115409 Russia *E-mail: sergeirubin@list.ru }

\begin{abstract}
{The formation mechanism of universes with distinctly different properties is considered within the framework of pure gravity in a space of D > 4 dimensions. The emergence of the Planck scale and its relationship to the inflaton mass are discussed.}


\end{abstract}
\maketitle

\section{INTRODUCTION}
The dynamics of our Universe is well described by a modern theory containing 30-40 parameters. The number of these parameters, whose values are determined experimentally, is too large for the theory to be considered final. In addition, it is well known that the range of admissible parameters must be extremely narrow (fine tuning of parameters) for the birth and existence of such complex structures as our Universe, which is difficult to explain. Extensive literature is devoted to a discussion of this problem \cite{Fine}. One way of solving the problem is based on the assumption about the multiplicity of universes with different properties  \cite{Landscape0},
\cite{Random}, \cite{Landscape1}. Rich possibilities for justifying this assumption are contained in the idea of multidimensionality of our space itself. The number of extra dimensions has long been a subject for debate. Thus, the Kaluza-Klein model originally contained one extra dimension. At present, for example, infinite-dimensional spaces \cite{Castro} and even variable-dimensional spaces \cite{Bleyer} are being discussed. In this paper, the concept of superspace is extended to a set of superspaces with different, unlimited from above numbers of dimensions. Based on the introduced extended superspace, we suggest the formation mechanism of universes with distinctly different properties and the emergence mechanism of the Planck scale. The probability of the quantum transitions that produce lower-dimensional subspaces is discussed.

Let us define the superspace  $\mathcal M_D =(M_D ;g_{ij})$ as a set
of metrics $g_{ij}$ in space $M_D$ to within diffeomorphisms. On a space-like section $\Sigma$, let us introduce a metric $h_{ij}$ (for details, see the book \cite{Misner} and the review \cite{Wilt}) and define the space of all Riemannian $(D - 1)$ metrics:
\begin{equation*}
Riem(\Sigma ) = \{ h_{ij} (x)\left| {x \in \Sigma } \right.\} .
\end{equation*}

The amplitude of the transition from one arbitrarily chosen section $\Sigma _{in}$ with the corresponding metric $h_{in}$ to another section  $\Sigma _{f}$ with a metric $h_f$ is
\begin{equation}
A_{f,in}=\left\langle {h_{f},\Sigma _{f}|{h_{in},\Sigma
}_{in}}\right\rangle =\int_{h_{in}}^{h_{f}}Dg\exp [iS(g)] .
\label{Amplitude}
\end{equation}
In what follows, we use the units $\hbar = c=1$. The topologies of the sections $\Sigma _{in}$ and $\Sigma _{f}$ can be different. We will be concerned with the quantum transitions in which the topology of the hypersurface $\Sigma _{f}$ is a direct product of the subspaces, $M_{D-1-d}\otimes M_d$. The space $M_d$ is assumed to be compact. Below, we will explore the question of what class of geometries on the hypersurface $\Sigma _{f}$ can initiate classical dynamics.

The entire analysis is performed within the framework of nonlinear gravity in a space of $D>4$ dimensions without including any matter fields. We discuss the emergence of the Planck scale and its relationship to the inflaton mass. The reduction in a lower-dimensional space is made in several steps to produce a cascade. Different cascades give rise to four-dimensional spaces with different effective theories and different numbers of extra dimensions.

The parameters of the low-energy theory turn out to depend on the topology of the extra spaces and vary over a wide range (see also \cite{Coleman88,Firo}), although the parameters of the original theory are fixed. This also applies to such fundamental concepts as, for example, the Planck mass and the topology of the extra space.
The absence of matter fields postulated here from the outset is a fundamental point. It is suggested that the metric tensor components for the extra (super)space at low energies will be interpreted as the matter fields in the spirit of theories like the

\section{THE SIMULTANEOUS FORMATION OF SPACE-TIME AND PARAMETERS OF THE THEORY}

Originally, the concept of superspace meant the set of various geometries \cite{Wheeler}; subsequently, the set of all possible topologies was included in it \cite{Misner}. Let us take the next step and extend the superspace to include spaces of various dimensionalities. To be more precise, let us define the extended superspace $\textrm{E}$ as a direct product of superspaces $\mathcal{M}$ of various dimensionalities:
\begin{equation}\label{ES}
\rm{E}=\mathcal{M}_{1}\mathcal{\otimes M}_{2}\mathcal{\otimes
M}_{3}\mathcal{\otimes}...\mathcal{\otimes M}_{D}...
\end{equation}
Here, $\mathcal{M}_{n}$ is a superspace of dimensionality $n = 1, 2, …,$ which is the set of all possible geometries (to within diffeomorphisms) and topologies.

Quantum fluctuations generate various geometries in each of the superspaces (spacetime foam) \cite{Wheeler,Misner}. The probability of the quantum birth of "long-lived" $3$-geometries and the conditions under which this occurs are discussed below in Section 4. In this section, we consider the corollaries of the hypothesis about the existence of an extended superspace.

Let us choose a space $M_D$ of some dimensionality $D$. Its topological structure can change under the effect of quantum fluctuations \cite{Misner, Konstantinov}. In particular, topologies are possible that admit of space foliations by space-like surfaces  $\Sigma$, as implied in Eq. (\ref{Amplitude}) for the transition amplitude. In what follows, we consider spaces that admit of a partition in the form
\begin{equation}
{M}_{D}= \mathbb{R}\otimes M^{(space)}_{D-1}, \label{timespace}
\end{equation}
where $\mathbb{R}$ represents the time-like direction.

Let us concretize the topology and metric on the space-like section  $\Sigma_{f}$ of amplitude (\ref{Amplitude}) by subjecting them to the following conditions:

i) the topology of the section $\Sigma_{f}$ has the form of a
direct product,
\begin{equation}
\Sigma_{f}={M}^{(space)}_{D-1}= M_{D_{1}}\otimes M_{d_{1}},
\label{twosubsp}
\end{equation}
where  ${D_{1}}$ and ${d_{1}}$ are the dimensionalities of the corresponding subspaces (in what follows, the compact subspace is denoted by $M_d, n = 1, 2, …$);

(ii) the condition for the curvatures of the subspaces ${D_{1}}$ and ${d_{1}}$ is satisfied:
\begin{equation}\label{Ineq0}
R_{D_{1}}(g_{ab})<<R_{d_{1}}(\gamma_{ij}),
\end{equation}

(iii) in the set of subspaces  $M_{d_{1}}$, we will choose the
maximum symmetric spaces with a constant curvature $R_{d_{1}}$, which is related to the curvature parameter $k$ in the standard way:
\begin{equation}\label{k}
R_{d_{1}}(\gamma_{ij})=kd_1 (d_1 -1).
\end{equation}
Otherwise, the topology and geometry of the subspace $M_{D}$ are arbitrary.

Let us choose the dynamical variables and the Lagrangian. We will write the metric of the space  $M_{D}$ as \cite{Carroll}
\begin{eqnarray}\label{interval}
ds^{2}&=&G_{AB}dX^{A}dX^{B}=g_{ac}(x)dx^{a}dx^{c}-b^{2}(x)\gamma_{ij}%
(y)dy^{i}dy^{j} \nonumber \\
&=&dt^2-h_{\alpha\beta}dx^{\alpha} dx^{\beta}-b^{2}(x)\gamma_{ij}%
(y)dy^{i}dy^{j}.
\end{eqnarray}
Here,  $g_{ac}$ is the metric of the subspace $R\otimes M_{D_{1}}$ with
signature $(+---...-)$, $b(x)$ is the radius of curvature of the compact subspace $M_{d_{1}}$, and  $\gamma_{ij}(y)$ is its positively defined metric. For a given foliation of the space by space-like surfaces, we can always choose the normal Gaussian coordinates, which is used in the last equality in (\ref{interval}).

The Einstein-Hilbert action for a gravitational field, linear in curvature R, completely describes the physical phenomena at low energies where gravity is important. However, it is clear that quantum effects inevitably lead to nonlinear corrections in the expression for the action \cite{Don}. In this case, the action ought to contain terms with higher derivatives in the form of polynomials of various degrees in Ricci scalar and other invariants. Thus, whatever the gravitational action we take as the basis, after applying the quantum corrections, it takes the form
\[
S = \int {d^N x\left( {R + \varepsilon _1 R^2  + \varepsilon _2 R^3
+ \varepsilon _3 R^4  + ... + \alpha _1 R_{AB} R^{AB}  + ....}
\right)}
\]
with a set of unknown coefficients dependent on the topology of the space \cite{Strominger,Rizzo,Cai}. However, the problem is not that acute, since the nonlinear (in Ricci scalar) theories can be reduced to the linear theory by a conformal transformation \cite{f(R)}. Moreover, in our papers \cite{Our}, we suggest a more general method for reducing arbitrary Lagrangians to the standard Einstein-Hilbert form in the low-energy limit. In this case, the problem of stabilizing the sizes of the extra dimensions \cite{Carroll} turns out to be quite solvable.

Thus, using theories with higher derivatives is inevitable when the quantum effects are taken into account. At low energies, such theories can provide the same predictions as general relativity, while having richer possibilities. For simplicity, we will restrict ourselves to an action quadratic in Ricci scalar:
\begin{equation}
S_{D}=\frac{1}{2}\int d^{D}X\sqrt{-G}\left[  R_{D}(G_{AB})+CR_{D}(G_{AB}%
)^{2}-2\Lambda\right]+ \int_{\partial M_D}Kd^{D-1}\Sigma \label{S1}%
\end{equation}
The contribution from the boundary  $\partial M_D$ is a term introduced by Hawking and Gibbons ($K$ is the spur of the second fundamental form of the boundary \cite{Gibbons}). In the subsequent analysis, the parameters $C$ and  $\Lambda$ are
supposed to be fixed and the entire variety of low-energy theories generated by action (\ref{S1}) arises owing to the various ways of reduction to lower-dimensional spaces.

Since the reasoning is given within the framework of a purely geometric approach, the scale in theory (\ref{S1}) is initially not fixed, while the parameters $C$ and $\Lambda$ are dimensionless. In the next section, we discuss the emergence of the Planck scale at low energies.

As was shown in \cite{Zhuk,Our}, theory (\ref{S1}) allows a compact stable space of extra dimensions with a nonzero vacuum energy density to be obtained. The observed low vacuum energy requires a very fine tuning of the parameter $C$ and $\Lambda$. Below, we suggest a mechanism that allows the parameters of the effective theory to be naturally varied over a wide range. In this case, the parameters are unique in each universe produced by a quantum fluctuation. This means that there exists a fraction of the universes with the required parameters and, hence, vacuum energy. Thus, the smallness of the dark energy density in our Universe means that the fraction of the universes similar to ours is small.

Let us return to the parameter variation mechanism. Owing to the special form of the chosen metric (\ref{interval}) (see \cite{Carroll}), the following relations hold:
\begin{eqnarray}\label{GRK}
&&  \sqrt{-G}=b^{d_{1}}\sqrt{-g}\sqrt{\gamma},\\
&&  R_{D}(G_{AB})=R_{D_{1}}(g_{ac})+b^{-2}R_{d_{1}}(\gamma_{ij})+K_b ,\\
&&  K_b\equiv-2d_{1}b^{-1}g^{ac}\nabla_{a}\nabla_{c}b-d_{1}(d_{1}-1)b^{-2}%
g^{ac}(\nabla_{a}b)(\nabla_{c}b) .
\end{eqnarray}
The covariant derivative $\nabla$ acts in the space $M_{D_1}$ . The
volume $V_{d_{1}}$ of the extra space $M_{d_{1}}$ depends on its topology, since it is expressed in terms of the intrinsic metric:
\begin{equation}
V_{d_{1}}=\int d^{d_{1}}y\sqrt{\gamma}.
\end{equation}
For convenience, let us introduce a scalar field
\begin{equation}\label{phi}
\phi(x)=b(x)^{-2}R_{d_{1}}(\gamma_{ij}),
\end{equation}
in terms of which the action takes the form
\begin{eqnarray}\label{S3}
&&  S_{D}=\frac{V_{d_{1}}}{2}\int d^{D_{1}}x\sqrt{-g}\left(
\frac{R_{d_{1}}}{\phi(x)}\right) ^{d_{1}/2}\cdot\\
&& \{R_{D_{1}}(g_{ab})+2C\phi(x)R_{D_{1}}(g_{ab})+CR_{D_{1}}(g_{ab}
)^{2}-2\Lambda+\phi(x)+C\phi(x)^{2}+ \nonumber \\
&&  2CK_b\left[ R_{D_{1}}(g_{ab})+\phi(x)\right]  +K_b +CK_b ^{2}\}.
\nonumber
\end{eqnarray}
Since the field  $\phi$ is uniquely related to the radius of curvature $b(x)$ of the compact space $M_{d_{1}}$, the existence of stationary solutions  $\phi(x)=\phi_m$ minimizing the action would mean the stability of the sizes of this space.

In the Jordan frame, according to (\ref{S3}), we have
\begin{eqnarray}
&&S_{D}=\frac{1}{2}\int d^{D_{1}}x\sqrt{-g}\{f(\phi)R_{D_{1}}(g_{ab}%
)+C_{J}(\phi)R_{D_{1}}(g_{ab})^{2}-2U_{J}(\phi)+K_{J}(\phi)],\label{S2}\\
&&f(\phi)=\mathcal{V}_{d_{1}}\phi(x)^{-d_{1}/2}\left[
1+2C\phi(x)\right]
\label{f}\\
&&U_{J}(\phi)=\mathcal{V}_{d_{1}}\phi(x)^{-d_{1}/2}\left[
\Lambda-\frac
{1}{2}\left(  \phi(x)+C\phi(x)^{2}\right)  \right] \label{La1}\\
&&C_{J}(\phi)=\mathcal{V}_{d_{1}}\phi(x)^{-d_{1}/2}C\label{c1}\\
&&K_{J}(\phi)=\mathcal{V}_{d_{1}}\phi(x)^{-d_{1}/2}\left[ K_b +2CK_b (R_{D_{1}%
}+\phi(x))+CK_b ^{2}\right] \label{K1}\\
&&\mathcal{V}_{d_{1}}  =V_{d_{1}}R_{d_{1}}^{d_{1}/2}%
\end{eqnarray}
Note the explicit dependence of the parameters of the derived effective theory on the topology of the extra space.
The mass  $m_{\phi}$ of a quantum of the scalar field  $\phi(x)$ is proportional to the second derivative of the potential at its minimum (in the Einstein frame) and can vary over a wide range. Note that the mass defined in this way is a dimensionless quantity. The following three situations are of greatest interest.

(1) There is no minimum of the potential, implying that the size of the extra space is nonstationary.

(2) A minimum of the potential exists and the condition
\begin{equation}
m_{\phi}^{2}\leq R_{D_{1}}(g_{ab}), \label{leq}%
\end{equation}
under which the scalar field evolves together with the metric of the "main" subspace $M_{D_{1}}$, is satisfied. It is this situation that is realized in inflationary models and that is discussed in the next section.

(3) A minimum of the potential exists and the condition
\begin{equation}
m_{\phi}^{2}\gg R_{D_{1}}(g_{ab}).\label{gg}%
\end{equation}
is satisfied. In this case, the scalar field  $\phi(x)$ rapidly relaxes to the potential minimum,
\begin{equation}
\phi(x)=\phi_{m}=Const,\label{bconst}%
\end{equation}
and does not change with time during low-energy processes. The third case is the most natural, since the relaxation time is proportional to the scale of the extra space $M_{d_{1}}$, which is small compared to the scale of the space $M_{D_{1}}$. Let us discuss this situation in more detail.

\newpage

\tablename{\hspace{0.2cm}1.%

  Dependence of the parameters $C_{D_{1}}$ and $\Lambda_{D_{1}}$ on the geometry of the extra space (factor $\mathcal{V}_{d_{1}}$).
The fixed parameters are: $D=D_1 + d_1 = 11, D_{1}
=4,\Lambda=-0.6,C=-1.9$. \newline }
$%
\begin{tabular}
[c]{|l|l|l|l|l|l|l|l|}\hline $\mathcal{V}_{d_{1}}$ & $10^{3}$ &
$10^{2}$ & $10$ & $1$ & $10^{-1}$ & $10^{-2}$ & $10^{-3}$\\\hline
$C_{D_{1}}$ & $\sim0$ & $\sim0$ & $\sim0$ & $-4\cdot10^{-13}$ &
$-4\cdot 10^{-6}$ & $-40$ & $-4\cdot10^{8}$\\\hline
$\Lambda_{D_{1}}$ & $-2\cdot10^{-6}$ & $-2\cdot10^{-5}$ &
$-2\cdot10^{-4}$ & $-2\cdot10^{-3}$ & $-0.016$ & $-0.16$ &
$-1.6$\\\hline
\end{tabular}
$

\bigskip

\tablename{\hspace{0.2cm}2.%

Dependence of the parameters  $C_{D_{1}}$ and $\Lambda_{D_{1}}$ on the dimensionality of the subspaces $M_{D_{1}}$ and $M_{d_{1}}$. The last row gives the radius of curvature $b_m$ for the extra space $M_{d_{1}}$. The fixed parameters are: $D=D_1 +
d_1=40,\mathcal{V}_{d_{1}}=100, \Lambda =-0.6,C=-1.9$. \newline}
\begin{tabular}
[c]{|c|c|c|c|c|c|}\hline
$D_{1}$ & $4$ & $10$ & $20$ & $30$ & $38$\\\hline
$C_{D_{1}}$ & $\sim0$ & $\sim0$ & $\sim0$ & $\sim0$ & $-3.5\cdot10^{-13}%
$\\\hline $\Lambda_{D_{1}}$ & $-8.7\cdot10^{-10}$ &
$-8.4\cdot10^{-4}$ & $-0.025$ & $-0.122$ & $-0.343$\\\hline $b_m$ &
$50.2$ & $44.5$ & $39.2$ & $32.8$ & $13.0$\\\hline
\end{tabular}

\bigskip

Assuming condition (\ref{bconst}) to be satisfied, let us perform a conformal transformation of the form (see, e.g., \cite{Bron})
\begin{eqnarray}
&&  g_{ab}=\left\vert f(\phi_m)\right\vert
^{-\frac{2}{D_{1}-2}}\tilde{g}_{ab},\label{conf}\\
&&  R_{D_{1}}=\left\vert f(\phi_m)\right\vert
^{\frac{2}{D_{1}-2}}\tilde {R}_{D_{1}},\nonumber\\
&&\sqrt{-g} =\left\vert f(\phi_m)\right\vert
^{-\frac{D_{1}}{D_{1}-2}} \sqrt{-\tilde{g}},\nonumber
\end{eqnarray}
Being applied to Eq. (\ref{S2}), it brings us back to the initial form of the action (in Eqs. \ref{La4}, the tildes were omitted for short)
\begin{eqnarray}\label{La4}
&&S_{D_{1}}  =\frac{1}{2}\int
d^{D_{1}}x\sqrt{-g}\{R_{D_{1}}(g_{ab})+C_{D_{1}
}R_{D_{1}}(g_{ab})^{2}-2\Lambda_{D_{1}}],\label{S4}\\
&&C_{D_{1}}  =\left\vert f\right\vert ^{\frac{4-D_{1}}{D_{1}
-2}}C_{J}(\phi_{m})=sign(f)\mathcal{V}_{d_{1}}\left\vert
f(\phi_{m})\right\vert
^{\frac{4-D_{1}}{D_{1}-2}}\phi_{m}^{-d_{1}/2}C,\label{c4}\nonumber \\
&&\Lambda_{D_{1}} =\left\vert f\right\vert ^{\frac{-D_{1}}{D_{1}
-2}}U_{J}(\phi)=sign(f)\mathcal{V}_{d_{1}}\left\vert
f(\phi_{m})\right\vert
^{\frac{-D_{1}}{D_{1}-2}}\phi_{m}^{-d_{1}/2}\left[
\Lambda-\frac{1}{2}\left(  \phi_{m}+C\phi_{m}^{2}\right) \right].
\nonumber
\end{eqnarray}
Recall that we consider the case where condition (\ref{bconst}) is satisfied, i.e., where the field $\phi$ is already at its minimum, $\phi =\phi_m$ and the kinetic terms were neglected.

Action (\ref{La4}) coincides in form with the initial one (\ref{S1}), but now in the subspace $M_{D_{1}}$ and with the renormalized parameters $C_{D_{1}}$ and  $\Lambda_{D_{1}}$ dependent on the volume  $V_{d_{1}}$ and curvature  $R_{d_{1}}$ of the extra space $M_{d_{1}}$. Tables 1 and 2 present the results of our numerical calculations. However, it is first necessary to note the following. The standard assumption that $V_{d}\sim L^{d}$ (where $L$ is the linear size of the space) is valid for "simple" spaces with a positive curvature (like $d$-dimensional spheres). For compact hyperbolic spaces, the situation is more interesting. The relationship between volume and linear size is defined by the asymptotic relation \cite{Kaloper}
\begin{equation}
V_{d_{1}}\simeq\exp\left[  \left(  d_{1}-1\right)  L/b_{m}\right]  ,L\gg b_{m}%
\end{equation}
where $b_m$ is the radius of curvature of the compact space corresponding to the value of the field $\phi$ at which the potential is at its minimum. Obviously, at a sufficiently large dimensionality of the extra space $d_1$, its volume can be large, while the linear size of the space can be
small. Thus, we can vary the parameter $\mathcal{V}_{d_{1}}$, along with
the parameters $C_{D_{1}}$ and  $\Lambda_{D_{1}}$, over a wide range without
coming into conflict with the experimental constraints on the size of the extra space. As an illustration, Tables 1 and 2 present the dependencies of the new parameters $C_{D_{1}}$ and $\Lambda_{D_{1}}$ on the topology of the compact extra
space $M_{d_1}$.

Thus, a theory similar to the original one that is valid in the space $M_{D}$, but with different parameters holds for the subspace $M_{D_{1}}$. An important thing is that although the original parameters $C$ and $\Lambda$ are fixed, the effective "secondary" parameters $C_{eff}=C_{D_{1}}$ and $\Lambda_{eff}=\Lambda_{D_{1}}$ of the reduced theory vary over a wide range. The specific values of the effective parameters depend on the random geometry and topology of the subspaces  $M_{d_1}, M_{D_1}$, produced by quantum fluctuations. The number of various topologies for a space of given dimensionality is at least countable. Consequently, the original theory with arbitrary, but fixed parameters (in our case, $C = -1.9, \Lambda = –0.6$) generates a countable set of reduced theories in lower-dimensional spaces $M_{D_1}$ that differ by the parameters. The interval of $C_{eff}$ and $\Lambda_{eff}$ can be
further extended if the next reduction of the subspace $M_{D_1}$ to an even smaller subspace $M_{D_2} \in M_{D_1}$ is taken into account. This is done in the next section

\section{CASCADE REDUCTION}

As was shown in the previous section, the reduction of the original theory to a lower-dimensional space generates a wide spectrum of secondary theories that differ by the parameters $C_{D_{1}}$ and $\Lambda_{D_{1}}$, which depend on the topology of the space produced by quantum fluctuations. The subspace $M_{D_{1}}$ on which the effective theory (\ref{S4}) is constructed, is also the subject to quantum fluctuations, which also lead to its partitions of the form
\begin{equation}
{M}_{D_1}=M_{D_{2}}\otimes M_{d_{2}} \label{twosubsp2}. \\
\end{equation}
The succeeding steps form a cascade:
\begin{equation}\label{Cascade}
M_{D_{1}}\rightarrow M_{D_{2}}\otimes M_{d_{2}};\quad
M_{D_{2}}\rightarrow M_{D_{3}}\otimes M_{d_{3}}\rightarrow...
\rightarrow M_{3}\otimes M_{d_{final}}.
\end{equation}
In this case, the parameters of the Lagrangian at the intermediate steps depend on the preceding steps of the cascade. Since the number of topologies at each step is at least countable, chain (\ref{Cascade}) rapidly branches out to ultimately produce an infinite set of effective theories that differ by the parameters. There is an infinite number of ways of "descending" from the initial space $M_D$ to the final one. The cascades of interest to us end with the formation of four-dimensional spaces $R\otimes M_{3}$ and compact extra spaces with a certain number of dimensions equal to $d_{final}$. Clearly, only a small fraction of the universes formed in this way are similar to our Universe.

Let us discuss the appearance of the Planck scale in this approach. Since so far we have dealt with purely geometric properties of spaces, introducing any scale seemed artificial. At the final step of cascade (\ref{Cascade}), our four-dimensional space emerges, while the metric tensor components for the extra space  $M_{d_{final}}$ are perceived by future observers as scalar and vector fields. Consequently, in contrast to the previous consideration, the kinetic terms $K_J$ in action (\ref{S2}) cannot be neglected. Using the low-energy limit (\ref{Ineq0}), we will neglect the highest powers of the scalar curvature $R_4$ and retain only the first terms in the expansion in powers of the derivatives $K_J =K(\varphi )\partial_{\mu} \varphi\partial \varphi^{\mu}+...$
Clearly, in this case, the most general form of the action within the framework of the Einstein frame is (see, e.g., \cite{Morris})
\begin{equation}\label{E-H_Action}
S\simeq \frac{V_{d_1}}{2}\int d^{4}x\sqrt{-g}[R_4 +
K(\varphi)(\partial \varphi)^2 -2V(\varphi)],
\end{equation}
Here the volume of the extra space $V_{d_1}$ enters this
expression explicitly. The form of the functions $K(\varphi)$ and $V(\varphi)$ depends on the topology and geometry of the extra space \cite{Our}. However, we will continue our analysis in general form.

The oscillations of the field $\varphi$ about its equilibrium position $\varphi
=\varphi_m$ are perceived by an observer as scalar field quanta. Near the potential minimum, $\min V(\varphi ) = V(\varphi _m ) = V_m$, action (\ref{E-H_Action}) is
\[
S \simeq \frac{{V_{d_1 } }}{2}\int {d^4 x\left[ {R_4 - K (\varphi _m
)(\partial \varphi )^2  - 2V (\varphi _m ) -  V'' (\varphi _m )
\left( {\varphi - \varphi _m } \right)^2 } \right]}
\]

The mass $m_{\varphi}$ of the scalar field quantum responsible for the inflation period (inflaton) is measured by an observer in dimensional units. At the same time, it is directly related to the form of the effective potential in the Einstein frame and for the standard form of the kinetic term. If the kinetic term $K(\phi)>0$ near the minimum of the potential $V(\phi)$, then a change of variables of the form
\begin{equation}\label{phichi}
x_{phys}  = x \cdot \frac{\sqrt{V'' (\varphi _m )}}{{m_\varphi \sqrt
{K (\varphi _m )} }};\;\varphi _{phys}  = (\varphi - \varphi _m )
\cdot m_\varphi \sqrt {V_{d_1 } } \frac{{K (\varphi _m )}}{\sqrt{V''
(\varphi _m )}}
\end{equation}
leads to a theory with the standard kinetic term:
\begin{eqnarray}\label{S5}
&&S = \frac{1}{2}\int {d^4 x_{phys} \left[ {V_{d_1 } m_\varphi ^2
\frac{K (\varphi _m )}{V'' (\varphi _m )} \cdot R_4 - (\partial
\varphi
_{phys} )^2  - m_\varphi ^2 \varphi _{phys}^2  - 2\Lambda } \right]} \\
&&\Lambda  \equiv V (\varphi _m )V_{d_1 } m_\varphi ^4 \frac{{K^2
(\varphi _m )}}{{V'' (\varphi _m ) ^2 }}\label{Lambda}.
\end{eqnarray}
If we denote
\begin{equation}\label{Mpl}
M_{pl}^2=V_{d_1 } m_\varphi ^2 \frac{K (\varphi _m )}{V'' (\varphi
_m ) } ,
\end{equation}
then we will obtain the standard form of the action for a scalar field $\chi$ in dimensional units:
\begin{eqnarray}\label{fin}
&& S=\int d^{4}x\sqrt{-g}\left[ \frac{M_{pl}^2}{2}R_4
+\frac{1}{2}(\partial \chi)^2 -U(\chi)\right] , \\
&& U(\chi )=\frac{1}{2}m_\varphi ^2 \chi^2  + \Lambda . \nonumber
\end{eqnarray}
In this case, the observed Planck mass is related to the parameters of the theory via Eq. (\ref{Mpl}).

Thus, three scales appear simultaneously at the final step of the cascade: the radius of curvature of the extra space $b_m$, the Planck mass $M_{pl}$ (see \ref{Mpl})), and the vacuum energy $\Lambda$ (see (\ref{Lambda})), which depend on the mass of the scalar field quantum $m_{\varphi}$. As we see, the Planck mass in the suggested approach is not a fundamental constant and depends on the final configuration of the cascade.

The cosmological constant $\Lambda \propto V(\varphi _m )$ depends on the effective parameters. The latter can vary over a wide range, as follows from the above discussion (see Tables 1 and 2). Hence, the cosmological constant in space $M_4$ also varies over a wide range, depending on the properties of cascade (\ref{Cascade}) that produced this space. Some of the cascades could lead to spaces $M_4$ with the observed value of the $\Lambda$ term. The extreme smallness of the latter means that only a small fraction of the cascades lead to the desired result. The parameters of the Universe are "fine tuned" when choosing the proper cascade.

Let us calculate the ratio of the Planck mass to the inflaton mass
\begin{equation}\label{Mplinf}
\frac{{M_{pl} }}{{m_\varphi }} = \sqrt {V_{d_1 } \frac{K(\varphi _m
)}{V''(\varphi _m )}}.
\end{equation}
Choosing the subspace  $M_{10}$  from set (\ref{ES}) and specifying the parameters $C=-1.9$ and $\Lambda =-0.6$ (see Tables 1 and 2), we will obtain for the dimensionality $d_1 = 6$ of the extra space
$$M_{pl}/m_{\varphi} \simeq 2\cdot 10^6$$%
This numerical value agrees well with the value that is commonly used in inflationary models.

\section{THE PROBABILITY OF THE BIRTH OF UNIVERSES WITH COMPACT EXTRA DIMENSIONS}

So far we have assumed that properties (\ref{timespace}) - (\ref{k}) are imposed on the structure of the space. In this section, we discuss the possibility that such topologies appear as a result of quantum fluctuations. Many papers are devoted to the quantum birth of the Universe. In our case, the situation is complicated by the fact that we consider nonlinear gravity and, in addition, there are extra dimensions whose stability should be taken into account specially. The birth of an n-dimensional space with extra dimensions within the framework of standard gravity was considered in \cite{Carugno,Ochiai}, where the stability regions of a compact subspace were also studied. The possibility of inflation in the presence of extra dimensions was explored in \cite{Carugno,Zhuk}. Quadratic (in Ricci scalar) gravity was investigated in this aspect in \cite{R2}.

The probabilities of the birth of the Universe calculated in different approaches differ radically from one another \cite{V1}. This may be indicative of both the imperfection of modern theories and the complexity of the subject. The ultimate goal of such calculations is to determine the probability of the appearance of a universe like ours. It would be unreasonable to expect this probability to be high, given that the parameters of the Universe are fine tuned. In this case, calculating the probability is of purely academic interest, because there are no causal relationships between the universes. At present, to justify the promising study, it is probably necessary and sufficient to prove that the fraction of the universes like ours is nonzero within the framework of a specific approach. In our case, this means that the probability of each transition (\ref{Cascade}) in the cascade is non-zero.

Classical trajectories on which the action is stationary make a major contribution to the transition amplitude (\ref{Amplitude}). Their shape depends on the boundary conditions and, in particular, on the properties of the manifold $\Sigma_{f}$. In our case, the metric on the hypersurface $\Sigma_{f}$ is defined by conditions (\ref{twosubsp}), (\ref{Ineq0}), (\ref{k}). Therefore, we will seek for classical trajectories that are subject to the same conditions on any section $\Sigma$ between the sections $\Sigma_{in}$ and $\Sigma_{f}$. The initial hypersurface $\Sigma_{in}$ can either be absent altogether (the Hartle-Hawking approach) or have a "zero geometry" (the interval between any two points of this hypersurface is zero in Vilenkin's approach). Below, we show that the transition probability depends weakly on the properties of the hypersurface $\Sigma_{in}$ in these cases.
As an example, let us consider the formation probability of the structure
\begin{equation}
\Sigma_f =M_{3}\otimes M_{d_{final}},\label{Sigmaf3}%
\end{equation}
that emerges at the last step of the cascade. Classical trajectories make a major contribution to the transition amplitude. The change of topology in the case of classical motion is unlikely. Therefore, classical trajectories consisting of hypersurfaces that also satisfy condition (\ref{Sigmaf3}) will be important for us. The topology of the D-dimensional Riemannian space between the sections $\Sigma_{in}$ and $\Sigma_{f}$ is then
\begin{equation}
\mathbb{R}\otimes M_{3}\otimes M_{d_{final}}.\label{decomp3}%
\end{equation}
As above (see (\ref{Ineq0})), the inequality  $R_{3} << R_{d_{final}}$, which allows the results of the previous section to be used, is assumed to be valid. Indeed, in the approximation used, action (\ref{S1}) transforms into a theory of form (\ref{E-H_Action}) and then into the habitual Einstein-Hilbert action (\ref{fin}).

Action (\ref{fin}) has been repeatedly used to study the quantum birth of the Universe (see, e.g., \cite{V1,HH,Linde,V2,Weiss,Fil}). However, in the papers devoted to the quantum birth of the Universe, the presence of a scalar field is usually postulated, while in our approach, this field constitutes the metric tensor components for the extra space. Therefore, we can use the results of numerous studies by briefly reproducing their main results.
The quantum birth of the Universe is generally studied within the framework of minisuperspace in which the interval is written as \cite{V1}
\begin{eqnarray}
&&ds^{2}=\sigma ^{2}\left[ N(t)^{2}dt^{2}-a(t)^{2}d\Omega
_{3}^{2}\right] ,\\
&&\sigma ^{2}=\frac{1}{12\pi ^{2}M^{2}_{Pl}}. \nonumber
\end{eqnarray}
(here, $N(t)$ is the laps function and $a(t)$ is the scale factor). The wave function $\psi (a)$ satisfies the Wheeler– DeWitt equation
\begin{equation}\label{WDW}
\left[ {\frac{{\partial ^2 }}{{\partial a^2 }} - W(a)} \right]\psi
(a) = 0,
\end{equation}
where the potential
\[
W(a) = a^2 (1 - H^2 a^2 ), a>0,\quad H=\frac{\sqrt{U(\chi )}}{6\pi
M^{2}_{Pl}}.
\]
The birth of the Universe is described by a tunnel transition with the forbidden region
\begin{equation}\label{ineq}
0<a<H^{-1}.
\end{equation}
The wave function in this region is \cite{V1}
\begin{equation}\label{psiT}
\psi(a)\simeq exp\left[{\int\limits_a^{H^{-1}}{\sqrt{-2W(a')}
da'}}\right] .
\end{equation}
The integral in this expression is ill-defined at the lower limit, where $a\rightarrow 0$. The approximation
$R_{3} << R_{d_{final}}$
does not work in this region, since $R_{3}=k/a^2
\rightarrow \infty$ and no explicit expression is defined for the potential. The same problem also takes place in other models for the quantum birth of the Universe \cite{Mijic}. Nevertheless, the integral calculated in this way has a meaning in the limit
\begin{equation}\label{H1}
H<<M_{Pl},
\end{equation}
when the region $a\sim 0$ is small compared to the entire domain of integration. Since our Universe was formed at $H\sim
10^{-6}M_{Pl}$, inequality (\ref{H1}) holds even at the inflationary phase. The conclusion that the result depends weakly on the behavior of the function near the singularity is also confirmed in the papers of other authors. Thus, for example, an initial wave function of the form $\delta(a-a_{in})$ was suggested in \cite{R2}, while the decay of a metastable vacuum from a state with a fixed energy was studied in \cite{Weiss,Fil}. In both cases, the initial conditions were shown to affect weakly the transition probability. The quantum birth of universes in multidimensional gravity is discussed in detail in \cite{Tavakol,Carugno}.

In Vilenkin's approach, the probability of the birth of the Universe is
$dP\propto
\exp\left[{\frac{+2}{3U(\chi)}}\right], $
while the Hartle-Hawking approximation yields $dP\propto
\exp\left[{\frac{-2}{3U(\chi)}}\right], $
Since the scalar field $\chi$ is uniquely related to the size of the extra space, the probability of the birth of extra dimensions depends on their linear size. For all their differences, the main thing in both approaches is that the event probability is nonzero and, hence, the fraction of the universes with given properties produced by a cascade of reductions is nonzero.

\section{DISCUSSION}

There exist several problems the solution of each is a serious task.

(1) The problems related to the extra dimensions - their number, compactification mechanism, and the possibility of experimentally verifying their existence.

(2) The problem of the nonlinearity of the gravitational action that inevitably arises as a result of quantum effects.

(3) The problem related to the fact that a Universe with a complex structure is formed at certain numerical values of the parameters of the theory. Generally speaking, the formulation of a future theory purporting to be the "ultimate" one should not contain any specific numerical values. Otherwise, an even more general theory explaining their origin will be required.

In this paper, we showed the interrelation between these problems. It turns out that their joint consideration facilitates rather than complicates the understand-ing of the ways of their solution. The cascade reduction process (\ref{Cascade}) allows at least a countable set of universes with various properties to be obtained. A Universe that is described by an effective low-energy theory with parameters whose values are determined by the specific cascade is formed at the final step of the cascade. The parameters vary over a wide range, indicating that the problem of fine tuning can be solved.

The idea of a cascade reduction developed here in the low-energy limit leads to the same results as the string theory, within the framework of which the concept of a "landscape" was introduced (see, e.g., \cite{Landscape0}, \cite{Landscape1}). The landscape is a complex form of a potential with many minima at which universes of various types exist. In contrast to the (super)string theory, the cascade reduction mechanism is based only on the assumption of space multidimensionality and action nonlinearity.

The Planck mass, the size of the extra dimensions, and the cosmological constant turn out to be dependent on the specific cascade and the method of determining the particle masses by an observer.
A necessary constituent of the cascade reduction is the possibility of the quantum birth of universes with compact extra dimensions. Here, we discussed the method of calculating the probability of this process within the framework of a gravitational action with higher derivatives.
Choosing the proper cascade that produced our Universe was not discussed here. Choosing the initial Lagrangian also remains an outstanding problem. As an example, we considered the simplest nonlinear (in curvature) Lagrangian. Using the approach suggested in \cite{Our}, our consideration can be easily generalized to more complex forms of the initial Lagrangian.

ACKNOWLEDGMENTS

I am grateful to K.A. Bronnikov, V.D. Ivashchuk,
M.I. Kalinin, and V.N. Melnikov for interest in the work and helpful discussions.

\end{document}